\documentclass[12pt]{article}
\usepackage{amssymb}
\usepackage{epsf}

\usepackage{epsfig}

\def\beq{\begin{equation}}
\def\eeq{\end{equation}}

\def\tev{\, {\rm TeV}}
\def\gev{\, {\rm GeV}}
\def\mev{\, {\rm MeV}}

\def\xfb{\, {\rm fb}}
\def\bea{\begin{eqnarray}}
\def\eea{\end{eqnarray}}
\def\beq{\begin{equation}}
\def\eeq{\end{equation}}
\def\mtilde{\tilde m}
\newcommand{\gsim}{\lower.7ex\hbox{$\;\stackrel{\textstyle>}{\sim}\;$}}
\newcommand{\lsim}{\lower.7ex\hbox{$\;\stackrel{\textstyle<}{\sim}\;$}}

\begin{document}

\setlength{\baselineskip}{0.25in}


\begin{titlepage}
\noindent
\begin{flushright}
MCTP-03-30\\
\end{flushright}
\vspace{1cm}

\begin{center}
  \begin{large}
    \begin{bf}

Implications of  Supersymmetry Breaking \\
with a Little Hierarchy between Gauginos and Scalars

    \end{bf}
  \end{large}
\end{center}
\vspace{0.2cm}
\begin{center}
James D. Wells \\
  \vspace{0.2cm}
  \begin{it}
Michigan Center for Theoretical Physics \\
University of Michigan, Ann Arbor, MI 48109-1120

  \end{it}

\end{center}

\begin{abstract}

From a theoretical point of view it is not hard to imagine gaugino
masses being much lighter than scalar masses.  The dominant
contributions to gaugino masses are then their anomaly-mediated
values.  Given current lower bounds on gauginos, which are near the
$W$ mass scale, considering a little hierarchy between weak-scale
gauginos and much heavier scalars requires suspending normal intuition
on finetuning and naturalness of the Higgs potential.  Nevertheless,
tantalizing perks come from the hypothesis: lessened flavor and CP
violation problems, more compatibility with gauge coupling unification
and third generation Yukawa unification, suppressed dimension-five
proton decay operators, and no problems satisfying the current Higgs
mass constraint for any value of $\tan\beta$ consistent with the top
and bottom Yukawa couplings remaining finite up to the grand unified
scale.  The Tevatron has little chance of finding any evidence of this
theory given current constraints.  The LHC does well looking for pair
production of gluinos which three-body decay into potentially
spectacular final states. Dark matter relic abundance can be
cosmologically interesting, but table-top experiments will not see
LSPs scattering off nucleons.  On the other hand, experiments looking
for monochromatic photons from LSP annihilations in the galactic halo
may find them.

\end{abstract}

\vspace{1cm}

\begin{flushleft}
hep-ph/0306127 \\
June 2003
\end{flushleft}

\end{titlepage}

\section{Introduction}

Of the various motivations for supersymmetry, perhaps none is more
powerful than its ability to stabilize disparate scales in a natural
way.  Unfortunately, the supersymmetric cure to the 
quadratic-sensitivity malady has
introduced several iatrogenic illnesses.  Results from flavor
changing neutral current experiments ($K-\bar K$ mixing, $\mu\to e\gamma$,
etc.), CP violation experiments (e.g., electric dipole moments of the neutron
and electron), and Higgs mass searches ($m_h>114\gev$ at 95\% C.L.)
all struggle to be consistent with low-scale supersymmetry.  One must make
additional assumptions about the superpartner spectrum, such
as the squarks must be degenerate and CP phases of superpartner
parameters ($\mu$, gaugino masses, $A$-terms) must be nearly zero.
Proton decay is another problem. Proposed solutions based on natural
R-parity arguments mollify dimension-four concerns, but dimension-five
operators still frighten the grand unified theory enthusiasts.

It is often assumed that superpartner masses need to be less than 
{\it about} $1\tev$ if the weak scale is not fine-tuned.  The problem is
that no one has a rigorous and defensible definition for ``about''.
If superpartner masses are at $10^8\tev$ most would agree that supersymmetry
would have little to do with stabilizing the weak scale to the quadratic
divergences of the Standard Model (SM) effective theory.  But what about
$3\tev$ or $48\tev$?  In the Yukawa coupling sector we have an apparent tuning
of $y_e/y_t=m_e/m_t\simeq 3\times 10^{-6}$.  If $\mtilde\simeq 150\tev$
we have a similar apparent finetuning in the electroweak sector
of $v^2/\mtilde^2\simeq y_e/y_t$.  This is not a complete argument, 
but rather a meek invitation to be less restrictive in thinking about what is
finetuned since an apparent finetuning might not be a real finetuning
once we learn more about the origin of Yukawa couplings and superpartner
masses.  

The primary phenomenological
motivation for considering a trans-TeV superpartner spectrum
is that it automatically solves most of the above-mentioned 
illnesses of supersymmetry.  Several approaches along these lines
have been presented in the literature, with heavy emphasis on
naturalness arguments.  Most emphasize heavy first two 
generation scalars~\cite{Dine:1990jd}-\cite{Cohen:1996vb}, while others
lift all three generation scalars~\cite{Feng:1999zg}.

In any case, electric dipole moments of the neutron and electron
are suppressed to well below experimental limits with trans-TeV squarks
and pseudo-scalar Higgs.
Kaon physics and B physics are identical to the Standard Model within current
experimental sensitivities if the squarks are sufficiently massive ---
tens of TeV for kaon physics, less for B physics.  

Another soothing effect of the little hierarchy of scalars to gauginos
is the suppression of proton decay in supersymmetric grand unified theories.
As the squark masses increase, the troubling dimension-five proton
decay operators are suppressed and proton decay is much less of
a concern~\cite{Arnowitt:1998uz}. In ordinary supergravity models with
a Bino lightest supersymmetric particle (LSP), the relic abundance increases
to unacceptable levels in much of parameter space when the scalar
masses are increased.  This is because the Bino annihilates most 
efficiently through $t$-channel sleptons, but when those masses are
too high the annihilation efficiency drops and the relic abundance
climbs very high such that the universe is matter dominated
too early.  We will see shortly that the supersymmetry spectrum
discussed in this letter gives the Wino the honor of being the
LSP.  The Wino LSP annihilates
very efficiently through ordinary gauge bosons and so the masses of
the scalars are mostly irrelevant to dark matter issues.  More will
be said about the dark matter situation in a subsequent section.

Yet another soothing effect of heavy scalars is the compatibility with
the Higgs boson mass constraint.  If all scalar superpartner
masses are above $5\tev$ the lightest Higgs mass boson is generally always 
above $114\gev$
for all values of $\tan\beta$ consistent with perturbative top and bottom
Yukawa couplings
up to the GUT scale. Thus, the interesting
low-$\tan\beta$ quasi-infrared fixed point region close to $\tan\beta \sim 2$
can be reincarnated without worrying about the Higgs mass.

The primary theoretical motivation for considering a trans-TeV superpartner
spectrum is the rather large hierarchy generated between superpartner
scalars and gauginos when there is no singlet to feel and transmit
supersymmetry breaking. Supersymmetry breaking can be parametrized
by a chiral supermultiplet $S=S+\sqrt{2}\psi\theta +F_S\theta^2$
whose non-zero $F_S$ component 
is the source of supersymmetry breaking.
Gaugino masses are generated via
\beq
\label{gaugino mass}
\int d^2\theta \frac{S}{M_{\rm Pl}}{\cal W}{\cal W}\sim 
 m_{3/2}\lambda\lambda
\eeq
where $m_{3/2}=\langle F_S\rangle/M_{\rm Pl}$.
The scalar masses are generated by
\beq
\label{scalar mass}
\int d^2\theta d^2\bar\theta \frac{S^\dagger S}{M_{\rm Pl}^2}\Phi^\dagger_i
\Phi_i \to m_{3/2}^2\phi^*_i\phi_i .
\eeq

If $S$ is charged (i.e., not a singlet), eq.~\ref{scalar mass} is unaffected,
whereas eq.~\ref{gaugino mass} is no longer gauge invariant.
(I am neglecting the grand unified theory possibility that a representation
of $S$ charged under
the unified group paired with that of the ${\rm Adj}^2$ contains a 
singlet~\cite{Anderson:1996bg}.) This is the generic expectation in
dynamical supersymmetry breaking where supersymmetry breaking order
parameters are charged and singlets are hard to come 
by~\cite{DSB singlets}.

In this case the leading-order contribution to the
gaugino mass is the anomaly-mediated 
value~\cite{Randall:1998uk,Giudice:1998xp},
\beq
M_\lambda=\frac{\beta(g_\lambda)}{g_\lambda}m_{3/2}
\eeq
where $\lambda$ labels the three SM gauge group.  The
gaugino masses are therefore one-loop 
suppressed compared to the persisting scalar mass
result of eq.~\ref{scalar mass}.  Some phenomenological implications
of this anomaly-mediated scenario with heavier squark masses was
presented in~\cite{Giudice:1998xp}.

Charged supersymmetry breaking therefore creates a one-loop hierarchy
between the gaugino masses (and $A$ terms)
and the scalar superpartners. With finetuning issues in play,
attempts have been made
to suppress the scalar masses via a specialized K\"ahler potential
of the no-scale variety.  Brane separations in extra dimensions
was one hope for this scenario~\cite{Randall:1998uk}, 
but it looks less likely when more complete models are 
constructed~\cite{Anisimov:2001zz}. String theory might
provide other ways to suppress
the scalar mass below 
$m_{3/2}$~\cite{Nelson:2002fk}.  Although ideas along these lines
might work out, the phenomenology presented
here is of the simplest version of the model where scalar masses are near the
gravitino mass.

\section{Superpartner Spectrum}

The numerical values of the light gaugino spectrum are
\bea
M_1 & \simeq & m_{3/2}/120 \\
M_2 & \simeq & m_{3/2}/360 \\
M_3 & \simeq & m_{3/2}/45.
\eea
As discussed above, 
the heavy superpartner spectrum of squark, slepton and sneutrino
masses $\tilde m_i$ should
have masses within factors of ${\cal O}(1)$ near the gravitino
mass $m_{3/2}$,
\beq
\tilde m_i \sim m_{3/2}~~({\rm scalar~masses})  
\eeq
unless there is a suppression from a special K\"ahler potential.

In minimal supergravity the region $m_0\gg m_{1/2}$ is the closest
match to the spectrum outlined above. It is well-known that radiative
electroweak symmetry breaking does not work well in this region.
The origin of the problem is universal soft masses.  When the Higgs and
top squark masses are all the same at the high scale, one needs a significant
$m_{1/2}$ to feed into the top squark mass to make it much heavier than
the Higgs soft masses at the low scale.  This drives $m_{H_u}^2$ negative
from loop effects, which enables the theory to 
satisfy the radiative electroweak symmetry
breaking constraint equation,
\beq
\label{ewsb constraint}
\frac{1}{2}m^2_Z+\mu^2=\frac{m^2_{H_d}-m_{H_u}^2\tan^2\beta}{\tan^2\beta-1}
\eeq
If $m_0\gg m_{1/2}$ one finds that $m_{1/2}$ effects are 
insignificant for the top squark, and
loop effects are not big enough to drive electroweak symmetry breaking.

However, universality is not needed in any way for the heavy scalar 
spectrum of supersymmetry considered here, and one can readily 
identify generic properties of the superpartner spectrum
that would give radiative electroweak symmetry breaking.  At the
high scale, if the soft masses satisfy,
\beq
m^2_{H_u}< m^2_{H_d} < m^2_{\tilde t},
\eeq
electroweak symmetry breaking normally will develop --- 
this ordering of masses makes it easier to satisfy eq.~\ref{ewsb constraint}.

The only problem left to have a full understanding of electroweak symmetry
breaking is the origin of the $\mu$ term.  Here, I offer 
no special insights into this recurring problem in
supersymmetric theories; however, there are many ideas one can
contemplate to explain the origin of the
$\mu$ term  in this context~\cite{Giudice:1998xp}. To be conservative
in the subsequent phenomenology discussion, I will assume that
$\mu\sim \tilde m_i$.  If $\mu\sim m_Z$, additional 
higgsino superpartner states
may be accessible, and detectibility would be enhanced in the discussion
that follows.

\section{Collider Searches}

For colliders, the most important implication is that gauginos are the
only superpartners accessible.  The neutral Wino $\tilde W^0$ 
is the lightest supersymmetric
partner (LSP), and its charged gauge partner $\tilde W^\pm$ is
very close by in mass.  The mass splitting in the limit of very massive
$\mu$ term is 
\beq
m_{\pi^\pm}< m_{\tilde W^\pm}-m_{\tilde W^0}< 165\mev
\eeq
as long as $m_{\tilde W}>80\gev$.  This is the maximally most challenging
mass splitting for experiment
because it is too large for $\tilde W^\pm$ to be
long-lived, which would generate a spectacular signature of a long, rigid
charged track decaying into a curling-up pion, and it is too small
to yield energetic and triggerable decay products $X$
in $\tilde W^\pm\to \tilde W^0+X$ decays.

Analysis have been done at LEP~\cite{lep chargino}, and the result is that 
$m_{\tilde W^\pm}> 88\gev$.  Searches at the Tevatron are more difficult
and focus on the production
of $\tilde W^+\tilde W^-\, +{\rm jet}$.  If one can
trigger on a jet in the events where its transverse energy is high
enough and then at the analysis stage look for
curled-track soft pions in the saved events, there is some hope of discovery.
However, it is unlikely that sensitivities will exceed that of
the LEP II results~\cite{Feng:1999fu}.

Analyses at the LHC should
focus on gluino production.  In minimal supergravity
the primary source of jets plus missing transverse energy is gluino production
in association with squarks.  However, there are no accessible squarks
here, and we therefore must ask how well
gluino pair production alone can probe gluino masses.  

\begin{figure}[bt]
\begin{center}
\includegraphics*[totalheight=3in]{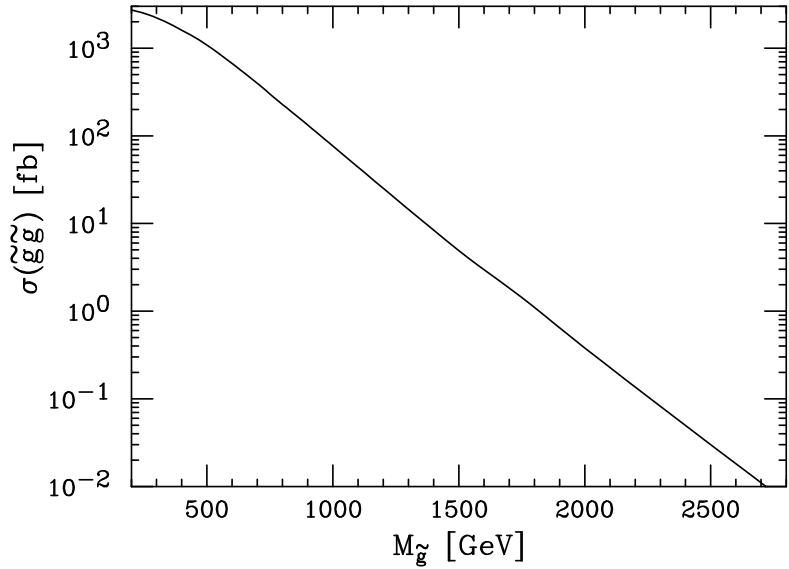}
\caption{Production cross-section of gluino pairs at the
$\sqrt{s}=14\tev$ LHC in the limit that all squarks
are much heavier than the gluino.  The final states
associated with gluino pair production are at least four jets
plus missing energy.}
\label{gluino production}
\end{center}
\end{figure}

Fig.~\ref{gluino production} shows the expected production cross-section
of gluino pairs at the $14\tev$ LHC 
in the applicable limit of much heavier squark 
masses~\cite{Baer:1999sp}. All events from $\tilde g\tilde g$ production
will produce at least four jets plus missing energy.  This is a 
spectacular signature in that draconian cuts on SM background can still
allow passage of the majority of the signal.  One LHC year at low luminosity
is expected to be about $10\xfb^{-1}$.  Accumulating $100\xfb^{-1}$
of data enables probes of the gluino up to an estimated mass scale
of $2\tev$.
Such a high gluino mass would correspond to a Wino mass of just over
$700\gev$ and scalar superpartner masses
of nearly $100\tev$.  Therefore, a $2\tev$ gluino search is certainly
more than impressive from a weak-scale naturalness point of view.

The decay branching fractions of the gluino are usually three body
through virtual squarks.  Two-body loop decays 
$\tilde g\to g\chi^0_i$~\cite{two-body}
are negligible in the limit that all squarks are equal in mass and
the $\mu$ term is large compared to the gaugino masss.  However, if
the top squarks are split and more than a factor of two lighter than
the other squarks one could get a sizeable $B(\tilde g\to g\chi^0_i)$
of several tens of percent.  In all cases, the branching fractions
are sensitive to the relative mass ordering of the squarks. 
For example, if the top squarks are lighter than other
squarks, as would be generally expected through renormalization group
running effects, one finds that gluino decays will produce a high-multiplicity
of top quarks.  This lends itself to discovery-friendly final states such
as 
\beq
pp\to \tilde g\tilde g\to tt\bar t\bar t \tilde W^0\tilde W^0~~{\rm and}~~
tt\bar b\bar b \tilde W^-\tilde W^-,~~{\rm etc.}
\eeq
The $\tilde W^-$ acts like a charged LSP in these events 
as it escapes all detection
because its SM decay products are too soft.  Both final states listed
above come with plenty of missing energy for further discrimination
from the Standard Model.

\section{Dark Matter}

In ordinary minimal supergravity the lightest superpartner is the 
bino, superpartner of the hypercharge gauge boson. The 
thermal relic abundance of this sparticle
is more or less compatible with the universe's cold dark matter needs
in much of the parameter space~\cite{Ellis:2003cw}.  
The price one pays
for this success is the gravitino and moduli problem.  The gravitino, which is
roughly the same mass as the LSP, decays during big bang nucleosynthesis
if its mass is less than a few TeV.
The gravitino and moduli must
be inflated away and not regenerated too copiously during the reheat
phase in this scenario. 

However, in our situation with anomaly-mediated gaugino masses, the Wino
is the LSP, which annihilates
and coannihilates very efficiently through SM gauge bosons.  The thermal
relic abundance is~\cite{Giudice:1998xp} 
\beq
\Omega_{\rm th}h^2\simeq 5\times 10^{-4}\left( \frac{M_2}{100\gev}\right)^2 
\eeq
and is cosmologically insignificant for weak-scale gauginos.  However,
if the Wino LSP mass is above about $2.5\tev$ one finds that the thermal
relic abundance is unacceptably high.  If all notions of finetuning
are rejected this would be perhaps the most important bound on
the entire superpartner spectrum in this scenario.

In this little hierarchy of scalars to gauginos, the 
gravitino mass is necessarily in the tens of TeV region.
Since it decays gravitationally, knowledge of its mass is all that
is needed to compute the decay lifetime.  The decays take place
after the putative
freeze-out time but before big bang nucleosynthesis.  Besides being
an interesting solution to the gravitino/moduli problem of weak-scale
supersymmetry, it also generates a non-thermal gravitino and moduli decay
source for relic LSPs~\cite{Gherghetta:1999sw}, which can naturally
account for all the cold dark matter.

Detecting Wino dark matter when squarks and the $\mu$-term are in the
tens of TeV is not possible with table-top detectors of LSP-nucleon
scattering.
The coherent scattering cross-section 
falls like $1/\mu^2$. In other words,
a
Higgsino component of the LSP 
is necessary to be sensitive to LSP-Nucleon
interactions, and since the LSP is nearly pure Wino the higgsino
component is not available for service.  
The spin-dependent contribution goes to zero as the $\mu$ term and
the sfermions
are very heavy and so they also do not contribute to scattering.
Therefore, the dark matter is invisible to 
table-top experiments.

However, Winos annihilate very efficiently and so one expects that all
experiments looking for  LSP annihilations in the galactic halo
would have an enhanced sensitivity.  For example, annihilations that 
produce $\bar p$'s and $e^+$'s 
are enhanced.  The annihilation channel that perhaps gains the most
if nature has Wino dark matter is the monochromatic 
two-photon final state~\cite{Wells:Pheno00,Ullio:2001qk}.  
The wino annihilation rate is even higher than the higgsino rate, which
is known to be large.  The 
cross-section for Winos annihilating into two photons~\cite{Bergstrom:1997fh} 
is a fairly constant 
value
\beq
2\sigma v(\gamma\gamma)
= (3-5)\times 10^{-27}\, {\rm cm}^{3}\, {\rm s}^{-1}
\eeq
for $m_{\tilde W^0}=0.1\tev - 1\tev$.

The virialized dark matter is moving at non-relativistic speeds of
only a few hundred kilometers per second, and so the photons that
result from this annihilation are monochromatic with energy
$E_\gamma=m_{\tilde W^0}$.  Under some astrophysical models developed
independently of dark-matter detection prospecting, the combination of 
GLAST and next generation Cerenkov detectors
may be able to see a signal for cross-sections above 
$3\times 10^{-27}\, {\rm cm}^{3}\, {\rm s}^{-1}$ for 
all dark matter masses between
$100\gev$ and $1\tev$~\cite{Ullio:2001qk}.  This bodes well
for detecting Wino LSPs. Another photon line from
annihilations into $Z\gamma$ might also be detectable at the energy
\beq
E_\gamma =m_{\tilde W^0}\left( 1-\frac{m_Z^2}{4m_{\tilde W^0}^2}\right)
~~({\rm from}~\tilde W^0\tilde W^0\to Z\gamma).
\eeq

Detecting a monochromatic photon line at the mass of the LSP would be
a tremendous aid to our understanding of what is happening
at the LHC.  In some circumstances at the LHC, mass differences of sparticles
can be measured to sub-GeV precision~\cite{Hinchliffe:1996iu}. 
However, measuring the overall
mass of a sparticle is difficult.  A monochromatic photon line from
LSP annihilations in the galactic halo would not only announce the
the discovery of dark matter, an important qualitative result, but it
would also give us
a very valuable number otherwise difficult to obtain: the mass of
the LSP.

\section{Conclusions}

There are other theoretical and phenomenological implications
of a little hierarchy between the gauginos and scalars.  
For example, recently it has been
understood that $b$-$\tau$-$t$ unification works well for very
large values of $\tan\beta\simeq 50$ but small finite $b$-quark mass
corrections~\cite{Blazek:2002ta}.  
This can be obtained naturally, from the Yukawa unification
point of view, if scalar superpartners are very heavy.  
Furthermore, gauge coupling unification in the supersymmetric standard model is
more exacting at the high-scale (i.e., less high-scale slop needed)
when the superpartners are in the multi-TeV region~\cite{Bagger:1995bw}, 
although when supersymmetry gets too high
(hundreds of TeV and above) 
the compatibility with gauge coupling unification fades.

The unification successes of supersymmetry, 
both gauge and Yukawa unification, are not diminished when the scalars
are very heavy. We have also seen that supersymmetric 
dark matter considerations do not cast dispersions on the idea either.
Only the naturalness of the electroweak symmetry breaking potential
apparently weakens with the increased scalar masses 
considered here.  Interestingly,
this aspect of supersymmetry 
is the least quantifiable success of supersymmetry, and it may
be that data will end up adjusting what we presume as natural.


\noindent
{\it Acknowledgments: }
This work was supported in part by the U.S.~Department of
Energy and the Alfred P. Sloan Foundation. 

\end{document}